\documentclass[conference]{IEEEtran}
\IEEEoverridecommandlockouts
\usepackage{cite}
\usepackage{amsmath,amssymb,amsfonts}
\usepackage{algorithmic}
\usepackage{graphicx}
\usepackage{textcomp}
\usepackage{xcolor}

\usepackage[ruled,linesnumbered]{algorithm2e}
\usepackage[many]{tcolorbox}

\newtcbtheorem{observation}{Observation}{
  colback=black!5,
  colframe=black!70}{d}

\def\BibTeX{{\rm B\kern-.05em{\sc i\kern-.025em b}\kern-.08em
    T\kern-.1667em\lower.7ex\hbox{E}\kern-.125emX}}
\begin{document}

\newcommand\dev[1]{\textcolor{black}{#1}}
\newcommand\commenta[1]{\textcolor{black}{#1}}
\newcommand\alek[1]{\textcolor{black}{#1}}

\newcommand\newk[1]{\textcolor{black}{#1}}

\title{Benchmarking of CPU-intensive Stream Data Processing in The Edge Computing Systems}

\author{\IEEEauthorblockN{1\textsuperscript{st} Tomasz Szydlo}
\IEEEauthorblockA{\textit{AGH University of Science and Technology, Poland} \\
\textit{and Newcastle University, UK}}
\and
\IEEEauthorblockN{2\textsuperscript{nd} Viacheslav Horbanov}
\IEEEauthorblockA{\textit{Newcastle University} \\
Newcastle upon Tyne, UK}
\and
\IEEEauthorblockN{3\textsuperscript{rd} Devki Nandan Jha}
\IEEEauthorblockA{\textit{Newcastle University} \\
Newcastle upon Tyne, UK}
\and
\IEEEauthorblockN{4\textsuperscript{th} Shashikant Ilager}
\IEEEauthorblockA{\textit{TU Wien} \\
Vienna, Austria}
\and
\IEEEauthorblockN{5\textsuperscript{th} Aleksander Slominski}
\IEEEauthorblockA{\textit{IBM TJ Watson} \\
Yorktown Heights, NY, USA}
\and
\IEEEauthorblockN{6\textsuperscript{th} Rajiv Ranjan}
\IEEEauthorblockA{\textit{Newcastle University} \\
Newcastle upon Tyne, UK}
}

%T. Szydlo, V. Horbanov, D. N. Jha, S. Ilager, A. Slominski, R. Ranjan

\maketitle

\begin{abstract}
Edge computing has emerged as a pivotal technology, offering significant advantages such as low latency, enhanced data security, and reduced reliance on centralized cloud infrastructure. These benefits are crucial for applications requiring real-time data processing or strict security measures. Despite these advantages, edge devices operating within edge clusters are often underutilized. This inefficiency is mainly due to the absence of a holistic performance profiling mechanism which can help dynamically adjust the desired system configuration for a given workload. Since edge computing environments involve a complex interplay between CPU frequency, power consumption, and application performance, a deeper understanding of these correlations is essential. By uncovering these relationships, it becomes possible to make informed decisions that enhance both computational efficiency and energy savings.
To address this gap, this paper evaluates the power consumption and performance characteristics of a single processing node within an edge cluster using a synthetic microbenchmark by varying the workload size and CPU frequency. The results show how an optimal measure can lead to optimized usage of edge resources, given both performance and power consumption.

% Edge computing has emerged as a pivotal technology with significant advantages such as low latency and improved data security. These benefits are crucial in applications requiring high security or real-time data analysis. However, edge devices \commenta{working in the edge clusters} are mostly underutilized, mainly due to the lack of a holistic performance profile that can help adjust the desired system configuration for a given workload. As there is a complex relationship between CPU frequency, power consumption, and application performance on \commenta{processing nodes}, understanding the correlation can uncover insights that can be utilized for better performance and energy efficiency.
% To address this gap, this paper evaluates the power consumption and performance of \commenta{the single processing node} %. The evaluation is performed on \emph{Raspberry Pi 4} \comment{single-board computer} 
% with synthetic 
% %\emph{stress-ng} 
% microbenchmark by varying the workload size and CPU frequency. The results show how an optimal measure can lead to optimized usage of edge resources, given both performance and power consumption.
\end{abstract}


\section{Introduction}
Edge devices are fundamentally transforming the execution landscape by enabling data processing closer to the source, thereby reducing latency and bandwidth usage while improving response times for real-time applications ~\cite{satyanarayanan2017emergence}. Despite their advantages, a significant challenge persists in managing power consumption effectively, as edge devices often operate under varying workloads, with overall system utilization remaining low in many scenarios~\cite{7842160, jiang2020energy}. %Typically, power consumption remains nearly constant regardless of workload intensity, leading to inefficiencies.

Data processing at the edge differs significantly from cloud-based processing in several key aspects. Unlike the cloud, where vast computational resources enable batch processing of large datasets, edge computing must handle real-time, continuous data streams generated by sensors. This means that edge systems process only a limited amount of data per unit of time and, even after completing one batch, must wait for new data to arrive before resuming computation ~\cite{renart2017data}. Additionally, cloud environments benefit from stable power availability, allowing for sustained high-performance processing, whereas edge devices operate under power constraints and often rely on battery or limited energy sources.

To maximize efficiency, edge devices must carefully balance performance and power consumption by dynamically managing their processing speed. One effective strategy is adjusting the CPU frequency, as there is a direct correlation between CPU frequency and power usage ~\cite{jiang2020energy}. Lowering the CPU frequency reduces energy consumption but may also degrade system performance, potentially leading to delays in real-time applications. Conversely, increasing the CPU frequency enhances processing speed but comes at the cost of higher power consumption. Given these constraints, understanding and optimizing this trade-off is essential for improving the overall efficiency, responsiveness, and sustainability of edge computing systems.

% \commenta{Data processing on the edge differs significantly from processing in the cloud environment. This is mainly because data from sensors is multimodal and usually consists of data streams. This means that the system must process a limited amount of information per unit of time, and even if it finishes processing previous data, it must wait for new ones to be processed ~\cite{renart2017data}. }  At the same time, edge devices should reduce their power consumption by managing their processing speed. One potential approach to this issue is to adjust the CPU frequency, as there is a direct relationship between CPU frequency and power consumption ~\cite{jiang2020energy}. Lowering the CPU frequency can reduce power consumption but also impact performance. Conversely, higher CPU frequencies can enhance performance but at the cost of increased power usage. Understanding and optimizing this trade-off is crucial for enhancing the efficiency of edge devices.

\begin{figure}
  \centering
  \includegraphics[width=0.75\columnwidth]{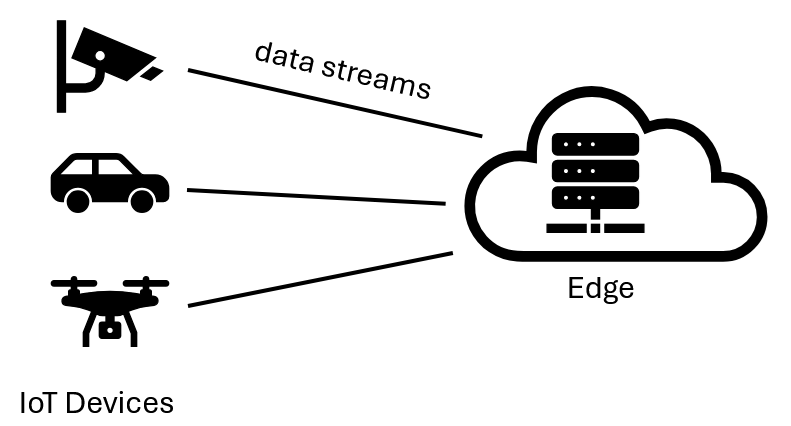}
  \caption{Stream data processing in the edge}
  \label{fig:idea}
\end{figure}

\begin{observation}{Data stream processing}{}
In the IoT-Edge-Cloud Continuum, edge devices should efficiently process the stream of data originating at the IoT sensors.
\end{observation}

Despite the importance of this issue, existing literature has not extensively addressed the \commenta{aspects of stream data processing}. %holistic evaluation of power consumption, CPU frequency, and workload status for edge devices. 
Most studies tend to focus on other aspects, such as performance optimization or power management for stream processing \cite{jha2019multiobjective,fragkoulis2024survey}.
This research provides a holistic evaluation of how dynamic CPU scaling influences both energy efficiency and computational responsiveness in edge environments. To achieve this, we propose a novel evaluation methodology that can be applied across various computing architectures, including x86, ARM, and RISC-V processing cores. Using a \emph{Raspberry Pi 4} as our testbed, we leverage \emph{stress-ng}, a powerful benchmarking tool, to simulate diverse workloads and analyze their impact on system behaviour.

% This paper aims to fill the gap by evaluating the interplay between power consumption, CPU frequency, and stream workload on edge devices. Our test environment comprises a \emph{Raspberry Pi 4}, and we propose %, a widely used platform in edge computing. 
% \commenta{evaluation methodology which might be applied for other computing modules with x86, ARM and RISC-V processing cores.} We employ \emph{stress-ng}, a versatile benchmarking tool, to simulate various workloads and measure their impact on system performance and power consumption.

Through systematic experimentation and performance-power trade-offs, we seek to identify optimal CPU frequency settings that balance power efficiency and performance across different workloads. 
% This research provides practical insights for developers and system designers, guiding them towards more efficient configurations for edge computing applications. 
Ultimately, our findings aim to contribute to the development of more sustainable and high-performance edge computing solutions.

The paper is organized as follows. Section \ref{sec:related} discusses the related work. In section \ref{sec:problem}, the research problem is defined. Section \ref{sec:methodology} discusses the methodology and testbed description while evaluation is discussed in Section \ref{sec:evaluation} accordingly. Finally, Section \ref{sec:conclusion} summarizes the results.

\section{Related work} \label{sec:related}

\subsection{Edge device benchmarking}
Benchmarking provides an effective way to understand a system's behaviour under various workload conditions \cite{howtobuildbenchmark_icpe}. While comprehensive benchmarking suites and results for cloud servers, such as SPEC \footnote{https://www.spec.org/cloud\_iaas2018/}, exist, there are limited benchmarking suites available for edge devices. Recently, there have been efforts to benchmark edge computing services for various workloads and hardware testbeds. For instance, EdgeBench \cite{EdgeBench} investigated the performance variance of different cloud service providers, such as Amazon AWS Greengrass and Azure IoT Edge. Similarly, MLPerf Edge Inference benchmarks and tinyML benchmarks~\cite{banbury2021mlperf} studied machine learning workloads on edge devices. This benchmark provides standard benchmarking suites and results measuring inference latency (samples/second) for popular pre-trained models on edge devices and accelerators, such as the NVIDIA Jetson AGX Orin. However, the effect of specific hardware configurations on energy and performance is not explored in detail. In this work, we specifically focus on benchmarking an exemplary edge device and investigating the effects on performance by varying the CPU frequency of the edge device.

\subsection{Power measurements}
Accurate measurement of the power consumption of edge devices is crucial in understanding the power consumption behaviour of edge resources under varied workload conditions. Traditionally, in cloud data centers, power measurements are carried out through onboard sensors equipped in each server. These sensors' data are exposed through software toolkits based on the Intelligent Platform Management Interface (IPMI). However, such standardized interfaces and platforms do not exist in edge devices. The lack of a standardized power monitoring framework in edge can be attributed to the high heterogeneity in edge devices and their software stack. Moreover, the demand for inexpensive edge devices makes designing sophisticated direct power measurements with onboard sensors impractical.

Software-based power measurement is an alternative and inexpensive method for power measurement. Such methods depend on analytical and predictive machine learning models developed based on empirical observations and profiling data on hardware testbeds. Some works have explored estimating energy consumption at the process level~\cite{SmartWatts, WattsKit}. Recently, tools such as Kepler have provided power consumption models for  Kubernetes container pods~\cite{kepler}. Kepler uses eBPF telemetry data and developed machine learning models to predict the power consumption of pods. Similar predictive models have been employed for measuring the energy consumption of edge devices such as Raspberry Pi~\cite{WattEdge}.

\begin{observation}{Power measurement}{}
Edge devices have limited capabilities in terms of their power consumption self-measurement.
\end{observation}

% \comment{intel vs ARM vs AMD}
% \comment{https://sustainable-computing.io/}
% \comment{https://github.com/sustainable-computing-io}
% \comment{Kepler/PEAKS/CLEVER}

\subsection{Governors in the Linux Kernel}
Since current processors can operate at different clock frequencies and voltage configurations, it is possible to adjust the speed of a CPU depending on the workload in the system. A processor can handle more instructions when the clock frequency is higher, but it comes with the cost of higher energy consumption. To balance power utilization, Linux Kernel offers the infrastructure referred to as CPU frequency scaling, or \textit{CPUFreq}, which provides several governors to meet the specific system need. A brief overview of each governor is given here\footnote{https://www.kernel.org/doc/Documentation/cpu-freq/governors.txt}:

\begin{enumerate}
    \item \textbf{\textit{performance}}: forces the CPU to run at the highest possible frequency.
    \item \textbf{\textit{powersave}}: forces the CPU to run at the lowest possible frequency.
    \item \textbf{\textit{userspace}}: allows to set the frequency manually.
    \item \textbf{\textit{ondemand}}: forces the CPU to run at the highest possible frequency when the workload is high, but sets the lowest possible frequency when the system load is low.
    \item \textbf{\textit{conservative}}: similar to the \textit{"on-demand"} governor, but it switches the frequency more gradually and is suitable for environments where sudden spikes in CPU frequency are undesirable.
    \item \textbf{\textit{schedutil}}: uses scheduler utilization data to make decisions about the CPU frequency.
\end{enumerate}

\newk{System configuration might also be overridden by the dedicated process and adjusted accordingly to the expected processing demand. We see this especially important for predictive management in context-aware applications where decisions are made based on factors other than system ones.}

\section{Problem statement} \label{sec:problem}
One of the main advantages of edge computing is the processing of sensor data close to the source of its creation. The source of such data may be IoT devices such as sensors, cameras and many others, as presented in Figure~\ref{fig:idea}. They generate a data stream that needs to be processed and sent further to the cloud. Unlike static data, where it is important to process it in the most optimal way possible, in the case of data streams, we cannot process more data than that generated by the devices. Therefore, the research problem is how to optimally configure the system to process only as much data as the size of the data stream.

\begin{figure}
  \centering
  \includegraphics[width=0.8\columnwidth]{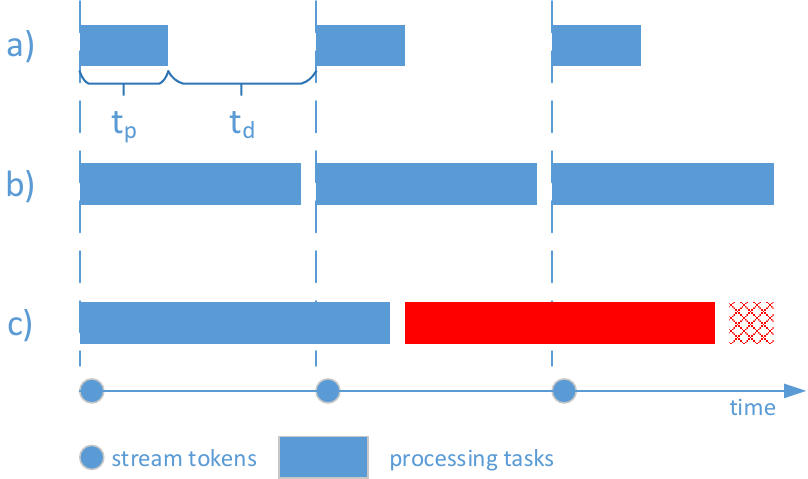}
  \caption{Stream data processing for various speed devices.}
  \label{fig:processing}
\end{figure}

\begin{observation}{Required processing speed}{}
The source of data defines the size of data stream to process - a slow stream of data requires less processing power than a multimodal one.
\end{observation}

Let's assume that the stream $s$ of sensor data requires $v_s$ synthetic operations per second to the processed. Without loss of the generality of the solution, we can assume that the stream of sensor data $s=(d,k)$ is composed of the tokens arriving in the time intervals $d$, and each token requires a particular number of operations $k$ to be processed. Therefore $v_s=k/d$. Tokens might represent images to be processed or time-windowed sensor data.

The processing of the data stream is shown in Figure~\ref{fig:processing} --- in case a) the computer's computing power exceeds the processing demand, and the data stream is processed without any delays; case b) indicates a situation where the machine's computing power is equal to the power needed to process the data stream; case c) is a borderline case, exceeding which results in a backlog of data for processing. The system's behaviour in such a situation is usually controlled by policy - input data is queued or skipped.

The processing speed $v_c$ of the edge node is reconfigurable, and its configuration is denoted as $c$. Therefore, the processing time of the stream tokens can be calculated as $t_p=k/v$, and the time delay for the next token to arrive is $t_d=d-t_p$. If the $t_d < 0$, then it means that the processing capacity of the node is insufficient to process the data stream.

In this research, we assume that the data stream processing algorithm is CPU-intensive. Therefore our problem of finding configuration $c$ for which processing stream $s$ is the most energy efficient could be defined as an optimisation problem:

\begin{align*}
    \textit{minimize} & \quad \quad P_c(s)=(P^{full}_c*t_p + P^{idle}_c*t_d)/d\\\\
    \textit{subject to} & \quad \quad c \\
    \textit{where} & \quad \quad t_d>0\\
     & \quad \quad t_p + t_d = d\\
\end{align*}

In the next section, the benchmarking methodology is discussed in more detail.

\section{Methodology} \label{sec:methodology}

%\begin{algorithm}
%\footnotesize
%\caption{Power benchmarking procedure}\label{alg:01_benchmarking_procedure}
%\begin{algorithmic}[1]
%
%\Require {$C$ - set of SUT configurations $\{c_1, ..., c_n\}$}
%
%\Procedure{Benchmark}{$C$}
%\State 
%\ForEach{$c_i \in C$} \Comment{Each configurations}
%    \State $r_e \gets 0$ \Comment{Total cost for environment $e_i$}
%    \ForEach{$m_r \in M^{e_i}$} 
%        \State $r_r \gets w_r * m_r(f, c)$ \Comment{Use model $m_r$}
%        \State $r_e \gets r_e + r_r$ \Comment{Add metric cost $r_r$}
%    \EndFor
%    \State $options(e_i) \gets r_e$
%\EndFor
%\\
%\State $r_{local} \gets 0$ \Comment{Estimate cost for IoT}
%\ForEach{$m_r \in M^{e_{local}}$} 
%    \State $r_r \gets w_r * m_r(f, c)$ \Comment{Use model $m_r$}
%    \State $r_{local} \gets r_{local} + r_r$ \Comment{Add metric cost $r_r$}
%\EndFor
%\State $options(e_{local}) \gets r_{local}$
%\\
%\State $e_{selected} \gets s(options)$ \Comment{Apply strategy $s$}
%\State \textbf{return $e_{selected}$}
%\EndProcedure

%\end{algorithmic}
%\end{algorithm}
This section discusses the test workload, testbed and system control parameters. The benchmarking procedure is presented as a pseudocode Algorithm~\ref{alg:procedure}. The steps of the procedure are discussed in the following subsections.

\begin{algorithm}[h]
    \caption{Benchmarking procedure}
    \label{alg:procedure}
%   \SetAlgoLined
    \KwIn{Set of SUT configurations $C$} 
    \KwOut{Power model of the SUT} 
    %\KwResult{$ID_{RN},~L_{RN},~Er_{RN}$ }

    $R = \{\}$

    \ForAll{configurations $c$ in $C$}{
        \ForAll{CPU limits $x$ in \{10,20,...,100\}}{
            $SUT \gets c$\;
            wait\;
            invoke $stress-ng$ with CPU limit $x$\;
            wait\;
            $p \gets $ power consumption from the power meter\;
            $m \gets $ metrics \textit{from stress-ng}\;
            $R \gets$ ($m$, $c$, $x$, $p$)\;
        }
    }
\end{algorithm}

% \subsection{Different tasks}
\subsection{Test Workload}
We leverage the Linux stress workload generator \textit{stress-ng}, which allows benchmarking the CPU and retrieving the performance summary as the result. Throughout the evaluation, there were primarily two types of \textit{stress-ng} calls used:

\begin{itemize}
    \item \texttt{stress-ng --cpu [c] --timeout [t]s --metrics-brief} \\
    used to stress test the CPU for a specified duration, utilizing a specified number of CPU stressors, and providing a summary of performance metrics after the test; \\
    
    \item \texttt{stress-ng --cpu [c] --cpu-load [u] --timeout [t]s --metrics-brief} \\
    allows setting the load percentage for the CPU stressors in a way that it stresses a specified number of CPU stressors at a given load percentage for a specified duration. \\
\end{itemize}

The parameter \texttt{c} specifies the number of cores, which is set to 0, enforces execution on all of the available cores. In the case of Raspberry Pi used in the experiments means execution on 4 cores;
%Here, \texttt{c} is the number of cores, specified to 0, which invokes all 4 cores of Raspberry Pi, during the experiment; 
\texttt{u} is the CPU utilization percentage (varies from 10\% to 100\% with a step of 10\%); \texttt{t} is the time in seconds, specified to 15s during the experiment.

\commenta{Because of the synthetic CPU benchmarking, we decided to rely on the embedded stress-ng metrics called \textit{bogo} operations per second. The size of the operations depends on the stressor being used, so we ensure during the evaluation that the same stressor and the tool version remain the same. }

\subsection{Frequency control}

The frequency control method depends on the system being evaluated. For the Raspberry Pi, to set a particular frequency of a processor, \newk{the \texttt{cpufreq-set} tool has been used with the following parameters:}

\texttt{sudo cpufreq-set -r -f [x]}

\noindent{}
where:

\begin{itemize}
%    \item \texttt{cpufreq-set}: The utility used to change CPU frequency settings.
    \item \texttt{-r}: applies the specified frequency to all CPU cores;
    \item \texttt{-f [x]}: specifies the desired CPU frequency. The value \texttt{[x]} should be given in kHz (kilohertz) or MHz (megahertz), depending on the CPU.
\end{itemize}

\subsection{Testbed description}

\begin{figure}
  %\centering
  \includegraphics[width=0.99\columnwidth]{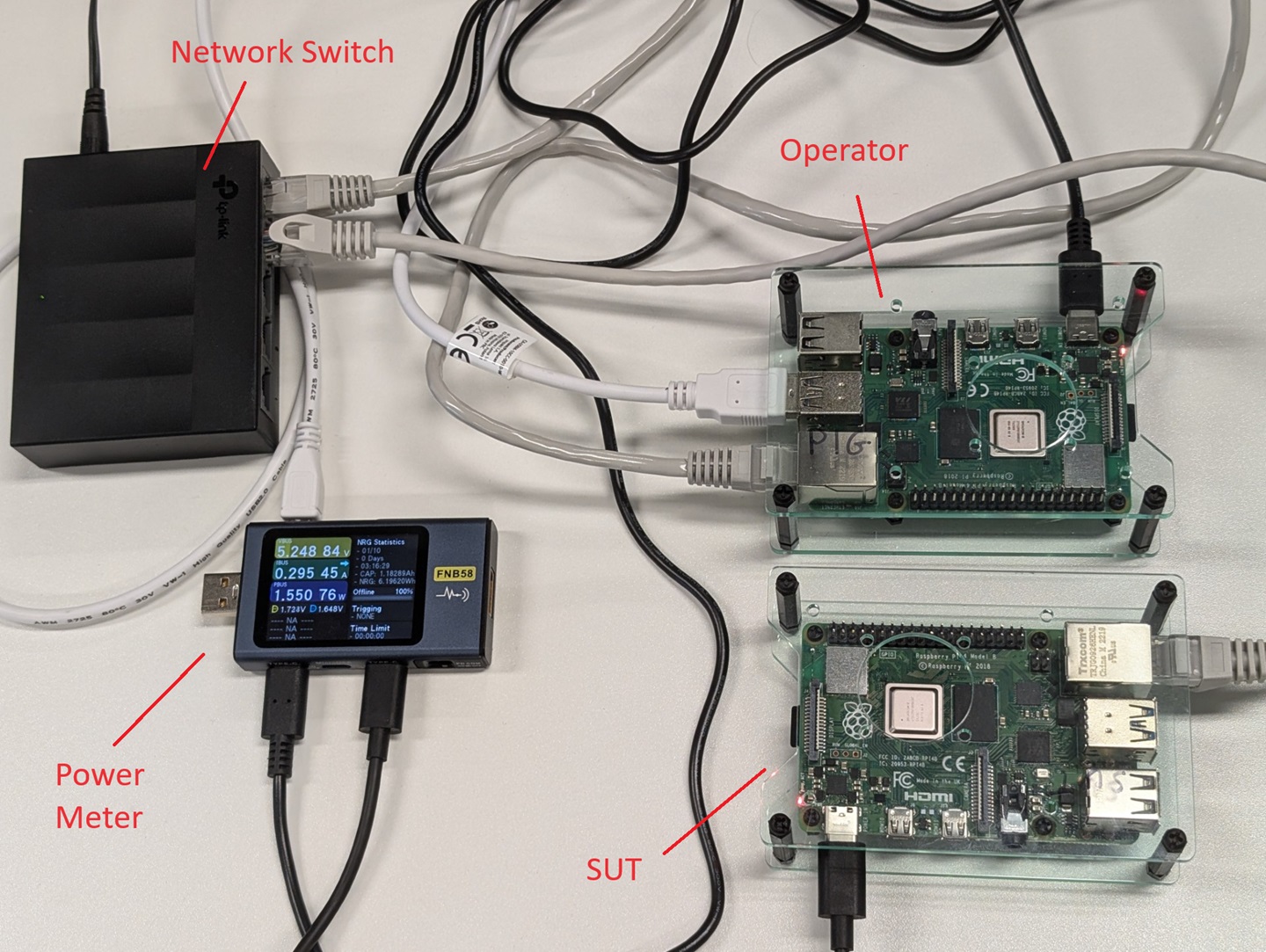}
  \caption{Testbed components}
  \label{fig:testbed_devices}
\end{figure}

The testbed consists of two \emph{Raspberry 4 Model B Rev. 1.5} \textit{(Broadcom BCM2711, Quad core Cortex-A72 (ARM v8) 64-bit SoC @ 1.8GHz)}, $4$ GB RAM (Figure~\ref{fig:testbed_devices}). The first RPi is used as the System Under Test (SUT) while the other
% , (where one is a system under the test, while another one 
is used for running an MQTT Broker, gathering the data and sending automated command pipelines. A power meter \emph{FNIRSI FNB58} is attached to the SUT to capture the power consumption. A network switch is used to provide an internal network and to connect the whole testbed to the Internet. An external computer is used to control the devices via SSH tunnel.
% power meter, network switch and an external computer to control the Raspberry Pi devices via SSH tunnel.

\subsection{Benchmark orchestration}

%\begin{figure}[!htbp]
%  \centering
%  \includegraphics[width=\columnwidth]{mqtt_governor.png}
%  \caption{Testbed architecture.}
%  \label{fig:testbed}
%\end{figure}

% \section{Testbed description}
% \subsection{Testbed description}

For evaluation purposes, we have developed a set of tools for benchmark execution\footnote{https://github.com/SlaviXG/mqtt\_system\_governor}.% as presented in Figure~\ref{fig:testbed}. 
The solution is based on the MQTT client-server setup where the server sends commands that get executed on each of the connected clients (subscribers), primarily used for measuring system performance and testing. The special feature of this application is the opportunity to build and run pipelines (sequences of terminal commands that get executed on client systems).

%\begin{figure}
%  \centering
%  \includegraphics[width=0.95\columnwidth]{setup_diagram.png}
%  \caption{IoT faults taxonomy.}
%  \label{fig:faults_taxonomy}
%\end{figure}

%The \textit{Commander} is a customizable MQTT client designed to send commands to specific clients (Systems Under Test, or SUTs) via the MQTT broker and receive feedback on command execution. It can be configured to format messages in JSON or plain text and extended or implemented in other custom programs to suit specific needs.

\section{Evaluation} \label{sec:evaluation}
In this section, our evaluation \commenta{aims to find the answers to the following questions:}%would like to answer the following questions:

\begin{enumerate}
    \item \textit{How does CPU frequency variation impacts the efficiency of edge devices?} \label{q1} This question tries to explore the relationship between CPU frequency and efficiency and determine whether the higher CPU frequency leads to higher efficiency or not.
    
    \item \emph{What is the relationship between CPU load and efficiency while varying the CPU frequencies?} \label{q2} This question aims to identify the frequency range where the CPU operates most efficiently for any given load.
    
    \item \emph{How does power consumption changes with increasing the CPU frequency for a given workload?} \label{q3} This question evaluates the relation between CPU frequency and power consumption for varying workload and assess the trade-off between performance and energy usage.
\end{enumerate}

% \subsection{Full CPU utilisation}
\subsection{CPU Frequency v/s CPU Efficiency} \label{CaseA}
\begin{figure}
  \centering
  \includegraphics[width=\columnwidth]{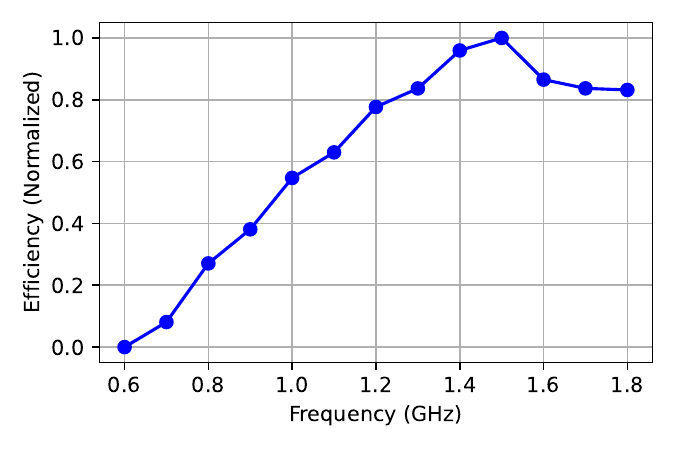}
  \caption{Result showing the CPU efficiency v/s CPU Frequency for the SUT.}
  \label{fig:full}
\end{figure}

To answer the question \ref{q1}, we evaluated the CPU efficiency of the SUT while varying the CPU frequency. Figure \ref{fig:full} shows the normalised CPU efficiency result which demonstrates a strict relationship between CPU efficiency and CPU frequency. As the frequency is increased from $600$ MHz to $1.5$ GHz, the CPU efficiency rises linearly. The linear increase in efficiency shows that within this range, the CPU can effectively utilize the additional clock cycles to perform more computations, therefore improves the overall efficiency. This is expected, as higher frequencies generally allow for more operations per second, which leads to better performance.

However, increasing the CPU frequency beyond $1.5$, GHz starts dropping the CPU efficiency slightly, as shown in Figure~\ref{fig:full}. This drop can be caused by increased power consumption and heat generation. %This shows that while increasing the CPU frequency can initially enhance the performance, there is an optimal point after which a further increase in the CPU frequency does not translate to efficiency gains and can even lead to a reduction in efficiency. 

\begin{observation}{CPU Frequency vs Efficiency}{}
Increasing the CPU frequency can initially enhance the performance, there is an optimal point after which a further increase in the CPU frequency does not translate to efficiency gains and can even lead to a reduction in efficiency.
\end{observation}

\begin{figure*}
  \centering
  \includegraphics[width=0.9\textwidth]{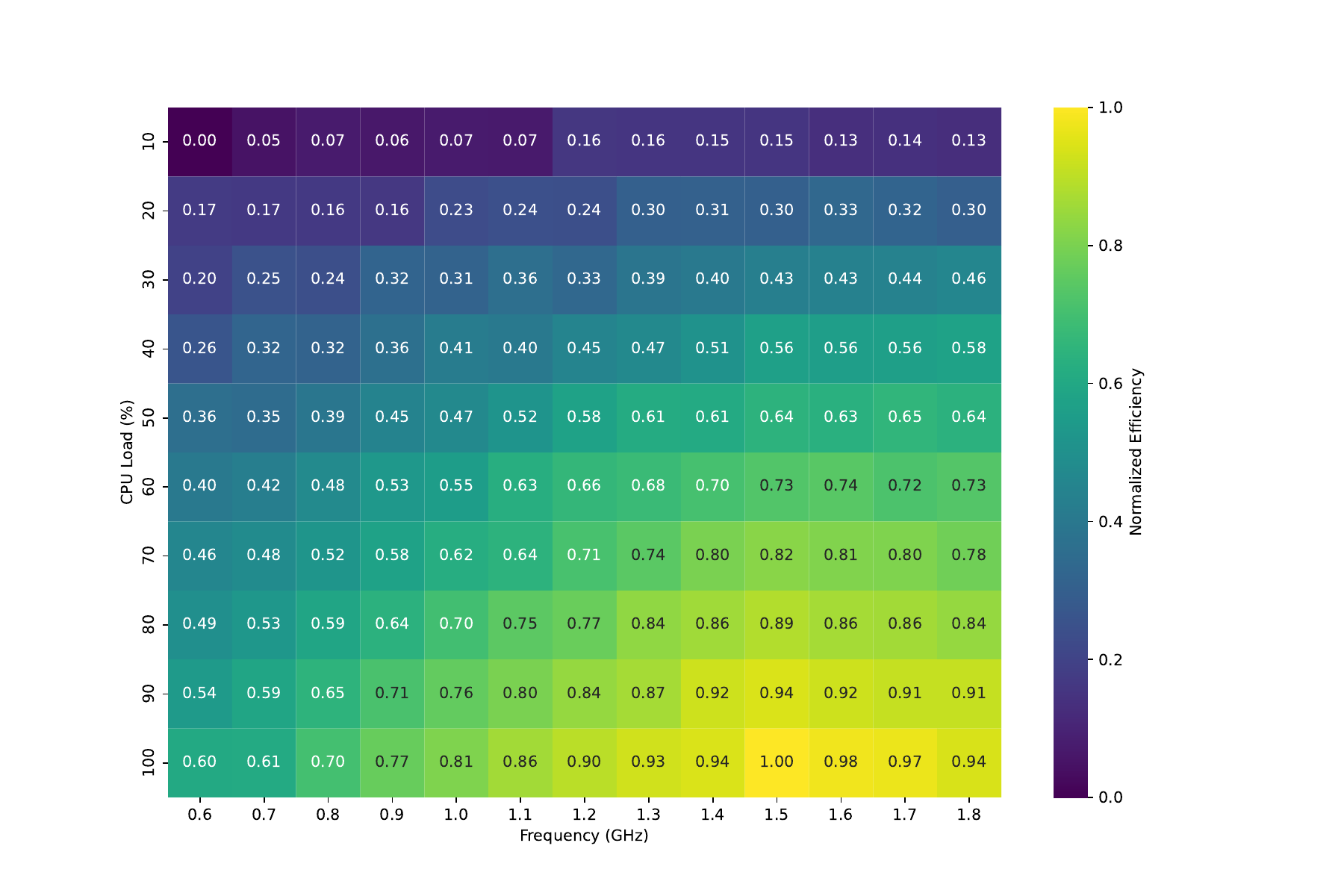}
  \caption{Result showing the variation of CPU efficiency by with varying CPU Frequency and CPU load for the SUT.}
  \label{fig:heat}
\end{figure*}

% The first experiment aimed to measure energy consumption for different SUT configurations. The results are shown in the Figure~\ref{fig:full}. The highest energy efficiency was achieved at 1.5GHz.

% \subsection{Benchmarking under variable load}
\subsection{CPU Load v/s CPU Frequency v/s CPU Efficiency} \label{CaseB}
Figure \ref{fig:heat} shows the evaluation result for CPU load and power efficiency across various CPU frequencies on the SUT. The result answers the question \ref{q2}. The heatmap shows that efficiency (normalized) varies significantly with changes in both CPU load and frequency. At lower frequencies, the efficiency remains relatively low for all CPU loads. This indicates that the CPU cannot effectively utilise its capacity at low frequencies, regardless of the workload, mostly due to the relatively high power consumption of the board peripherals.

As the frequency increases, the efficiency increases gradually. This trend is easily seen in the mid to high frequency range (around $1.2$ GHz to $1.5$ GHz), where the efficiency reaches maximum. The CPU can handle higher loads at these frequencies more efficiently, which can be attributed to the improved balance between load and efficiency. The maximum efficiency observed at $1.5$ GHz aligns with the previous result, further confirming that it is optimal for balancing the performance of the SUT.

Beyond 1.5 GHz, the efficiency starts to saturate and even slightly decline at the highest frequency of $1.8$ GHz, particularly at maximum CPU loads. This decline can be explained by the increased power draw and thermal output at these higher frequencies, as explained in the result \ref{CaseA}.%, which can cause the CPU to throttle and reduce its effective processing power.

\begin{observation}{Power-efficient CPU utilisation}{}
CPU-intensive processing is the most power-efficient when the CPU is almost fully utilised.
\end{observation}

% The second experiment aimed to measure energy consumption for various SUT operating parameters. The experiment scenario is described by the algorithm, and then the results are collected and presented on a graph in Figure~\ref{fig:heat}. The data has been normalized and coincides with the previous chart for 100% load.

\subsection{Power consumption v/s CPU Frequency}

\begin{figure}
  \centering
  \includegraphics[width=0.99\columnwidth]{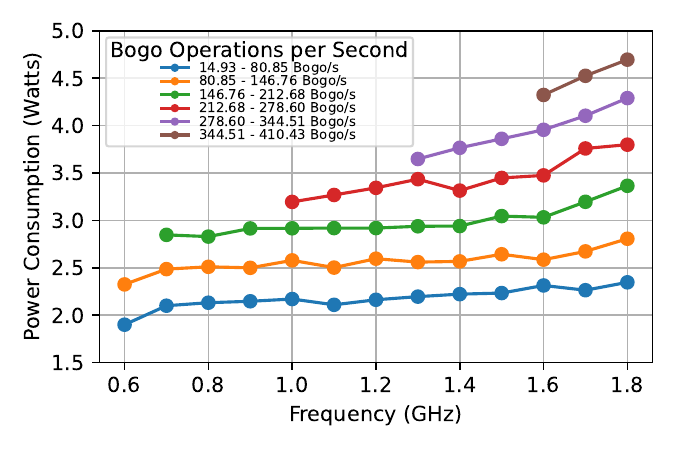}
  \caption{Result showing the Power consumption v/s CPU Frequency.}
  \label{fig:streams}
\end{figure}

The third result, as presented in Figure~\ref{fig:streams}, shows the power consumption of the SUT as a function of CPU frequency with the varying \textit{bogo} operations per seconds. The result shows a clear trend where power consumption increases as the CPU frequency rises for a given \textit{bogo} operations for seconds. At lower frequencies ($600$ MHz to $1.2$ GHz), the power consumption is relatively low with $2.0$ to $2.5$ watts, which reflects the reduced energy demands at these operational speeds.

As the frequency increases beyond 1.2 GHz, there is a more pronounced rise in power consumption, which sharply reaches up to $1.8$ GHz. At this frequency, power consumption reaches approximately $4.75$ watts. This significant increase is due to the CPU requiring more power to sustain higher operational speeds, enabling it to perform more \textit{bogo} operations per second. The higher power consumption at high CPU frequencies is also consistent with the CPU's increased need for energy to maintain performance and manage heat dissipation.

This data shows the trade-off between performance and power efficiency. While higher frequencies enable more operations per second, thereby boosting performance but they also lead to substantially higher power consumption. This relationship highlights the necessity of optimising CPU frequency for specific applications to balance performance demands with energy efficiency, especially in scenarios where power consumption is a critical factor.

\begin{observation}{Power-efficient frequency control}{}
For a given data stream, the lower the CPU frequency, the lower the energy consumption. Therefore, frequency scheduler should keep high CPU utilisation.
\end{observation}

%\subsection{Image classification models}

%Lorem ipsum...

%\begin{figure}
%    \centering
%    \includegraphics[width=0.95\linewidth]{rpi5_bench.png}
%    \caption{Enter Caption}
%    \label{fig:enter-label}
%\end{figure}

%\begin{observation}{External NN coprocessor}{}
%Inferencing using an external NN accelerator only slightly impacts the main CPU of the device.
%\end{observation}

\section{Conclusions} \label{sec:conclusion}
The results of our research on synthetic workload indicate that processing data streams in the edge-cloud continuum imposes specific requirements on the way data is processed. Unlike batch jobs, data streams must be processed on an ongoing basis. Otherwise, the amount of unprocessed data will increase, causing data stream fluctuations and deterioration of the quality of data processing. However, what is important is the fact that the amount of data to be processed is determined by the source of sensory data.

Our results indicate that for CPU-intensive tasks, optimal energy efficiency is achieved with high CPU utilization. Therefore, the processor frequency adaptation strategy should maintain a high level of processor utilization by decreasing CPU frequency in such a way to achieve its utilisation close to 100\%.
The indicated observations will be used in further work to intelligently manage an edge cluster based on lightweight Kubernetes distributions such as \textit{k3s} or \textit{MicroShift} \alek{eg. composed of multiple versions of Raspberry Pi such as \textit{Raspberry Pi Zero} and other architectures such as \textit{RISC-V}}. The energy model of edge processing nodes will be applied to the CNCF Kepler project and, together with Prometheus metrics, will be used for holistic resource scheduling.

\commenta{Another area of research will be analyzing the energy efficiency of ML video stream processing for various neural network accelerators, such as mobile GPUs and dedicated chipsets, including Hailo and Kinara. This will be used for smart scaling of serverless machine learning model serving based on the \textit{Knative} framework.}

%\section*{Acknowledgments}
%This study was supported by the UKRI EPSRC EP/Y028813/1 ”National Edge AI Hub for Real Data: Edge Intelligence for Cyberdisturbances and Data Quality” and European Union HORIZON-MSCA-2021-SE-01-01 project Cloudstars (g.a. 1010862).

\section*{Acknowledgment}
This study was supported by the European Union HORIZON-MSCA-2021-SE-01-01 project Cloudstars (g.a. 1010862) and the UKRI EPSRC EP/Y028813/1 ”National Edge AI Hub for Real Data: Edge Intelligence for Cyberdisturbances and Data Quality”.

\bibliographystyle{IEEEtran}
\bibliography{sample-bibliography} 

%\begin{thebibliography}{00}
%\bibitem{b1} G. Eason, B. Noble, and I. N. Sneddon, ``On certain integrals of Lipschitz-Hankel type involving products of Bessel functions,'' Phil. Trans. Roy. Soc. London, vol. A247, pp. 529--551, April 1955.
%\bibitem{b2} J. Clerk Maxwell, A Treatise on Electricity and Magnetism, 3rd ed., vol. 2. Oxford: Clarendon, 1892, pp.68--73.
%\bibitem{b3} I. S. Jacobs and C. P. Bean, ``Fine particles, thin films and exchange anisotropy,'' in Magnetism, vol. III, G. T. Rado and H. Suhl, Eds. New York: Academic, 1963, pp. 271--350.
%\bibitem{b4} K. Elissa, ``Title of paper if known,'' unpublished.
%\bibitem{b5} R. Nicole, ``Title of paper with only first word capitalized,'' J. Name Stand. Abbrev., in press.
%\bibitem{b6} Y. Yorozu, M. Hirano, K. Oka, and Y. Tagawa, ``Electron spectroscopy studies on magneto-optical media and plastic substrate interface,'' IEEE Transl. J. Magn. Japan, vol. 2, pp. 740--741, August 1987 [Digests 9th Annual Conf. Magnetics Japan, p. 301, 1982].
%\bibitem{b7} M. Young, The Technical Writer's Handbook. Mill Valley, CA: University Science, 1989.
%\end{thebibliography}

\end{document}